# 2bNIRS: a portable, multi-distance, broadband oximeter and cytochrome-c-oxidase monitoring system for in vivo applications

L. Giannoni[a]*, A. Barraclough[a], C. Carnati[b], F. Lange[a] and I. Tachtsidis[a]

[a] *Department of Medical Physics and Biomedical Engineering, University College London, London, UK*
[b] *Dipartimento di Elettronica, Informazione e Bioingegneria, Politecnico di Milano, Milan, Italy*

**Corresponding author:* l.giannoni@ucl.ac.uk

**Abstract:** Conventional near-infrared spectroscopy (NIRS) instruments typically employ 2-3 wavelengths for monitoring tissue haemodynamics. However, the use of broadband illumination (hundreds of wavelengths) unlocks also the targeting of tissue metabolism by exploiting the wide differential-redox peak of absorption of cytochrome-c-oxidase (CCO). For this purpose, we present here a novel, broadband (780-900nm), multi-distance NIRS system, called 2bNIRS. Compact and portable ($20 \times 30 \times 40$ cm$^3$), 2bNIRS was validated using a dynamic, optical phantom containing blood and yeast, and benchmarked against a multi-wavelength time-domain NIRS (TD-NIRS) system. Validation confirmed its accuracy in tracking haemodynamic and metabolic changes, with superior oximetry performance achieved via our BRUNO algorithm, which integrates broadband spectral fitting with spatially-resolved analysis. Preliminary *in vivo*, human applications demonstrated 2bNIRS applicability to monitoring both muscle tissue, during arm cuff occlusions, and the brain, during frontal cortex activation.

**Introduction:** Optical spectroscopy techniques, such as near-infrared spectroscopy (NIRS) and diffuse optical tomography (DOT), has the benefit of providing non-invasive, low-cost and flexible continuous monitoring of tissue haemodynamics and oxygenation at bedside, as well as in various naturalistic environments [1, 2]. These methodologies typically consist of emitting and detecting 2-3 wavelengths of NIR light diffused through tissue, such as muscle and the brain. This approach enables the estimation of changes in oxyhaemoglobin ($HbO_2$) and deoxyhaemoglobin (HHb), by exploiting their distinct absorption characteristics within the NIR range, where tissue transparency allows for deeper light penetration. Furthermore, when more than one pair of source-to-detector distances are used, continuous-wave, optical spectroscopy methodologies unlock the ability to obtain absolute tissue oximetry via multi-distance algorithms, which is of enormous significance in the clinics. The majority of these algorithms relies on the estimation of the slope of the light attenuation with distance to quantify absolute tissue saturation ($StO_2$), for example spatially-resolved spectroscopy (SRS) methods [3]. More advanced methods, such as dual slope spectroscopy (DSS), exploit the use of more than two sources and detectors in symmetric configuration to enhance depth sensitivity to deeper layers [4].
Recently, multi-wavelength (more than 6 wavelengths) or broadband (hundreds of wavelengths) NIRS (bNIRS) approaches exploit the richness of the spectral information to target additional chromophores beyond haemoglobin. bNIRS, in particular, has demonstrated its successful capability to target tissue metabolism via the measurement of the optical signature (in the range 780-900 nm) of the redox states of cytochrome-c-oxidase (CCO), an enzyme in the electron transport chain of the mitochondria involved in more than 95% of adenosine triphosphate (ATP) production [5]. Via bNIRS, absolute oximetry can also be performed using broadband fitting algorithms for absolute quantification of tissue saturation. Furthermore, Kovacsova *et al.* recently merged the advantages of bNIRS and multi-distance spectroscopy to produce an algorithm, called BRUNO, that combines broadband fitting with SRS to further improve the accuracy in absolute tissue oximetry [6]. Therefore, having an instrument that combines the potentials of bNIRS and multi-distance spectroscopy to continuously monitor changes in $HbO_2$, HHb and oxidised CCO (oxCCO), as biomarkers of haemodynamics and metabolism, respectively, as well as absolute tissue oximetry, could be of extreme significance and beneficial impact in several applications, particularly clinical ones, such as in neonatal intensive care [7]. For this kind of purposes, where space is limited and ergonomics is crucial, miniaturising and making devices portable and easily integrable in crammed environments (such as a hospital room) is crucial. Such unmet need was the underlying motivation for developing our 2bNIRS system, a compact and fully transportable broadband oximeter and CCO monitor, allowing users to access livestream information on haemodynamics, metabolism and absolute oxygenation with ease and high versatility. 2bNIRS, here presented and validated with some exemplary use-studies, could represent the ultimate clinical tool for a rapid and comprehensive insight into both oxygen delivery and consumption in tissues.

**Material and methods:** The 2bNIRS system (Fig. 1) is a novel custom-made device, fully portable and compactly fitted in a box of 20 x 30 x 40 $cm^3$. It utilises two broadband, tungsten-halogen sources (Ocean Insight, HL-2000-HP), and two spectrometers (Wasatch Photonics, WP-785XM-R-IS-25-OEM+NF) working in the range 780 to 900 nm, at 0.25-nm resolution. Fibres bundles conduct light via a tailored, 3D-printed optode probe, that enables adaptability and customisation to different geometries and curvatures of the targets, as well as offering dual source-to-detector distances of 2 and 3 cm. This allows the users to employ all the main multi-distance methodologies to estimate absolute tissue oxygenation, including SRS [3], Dual Slope [4] and the BRUNO algorithm [6].

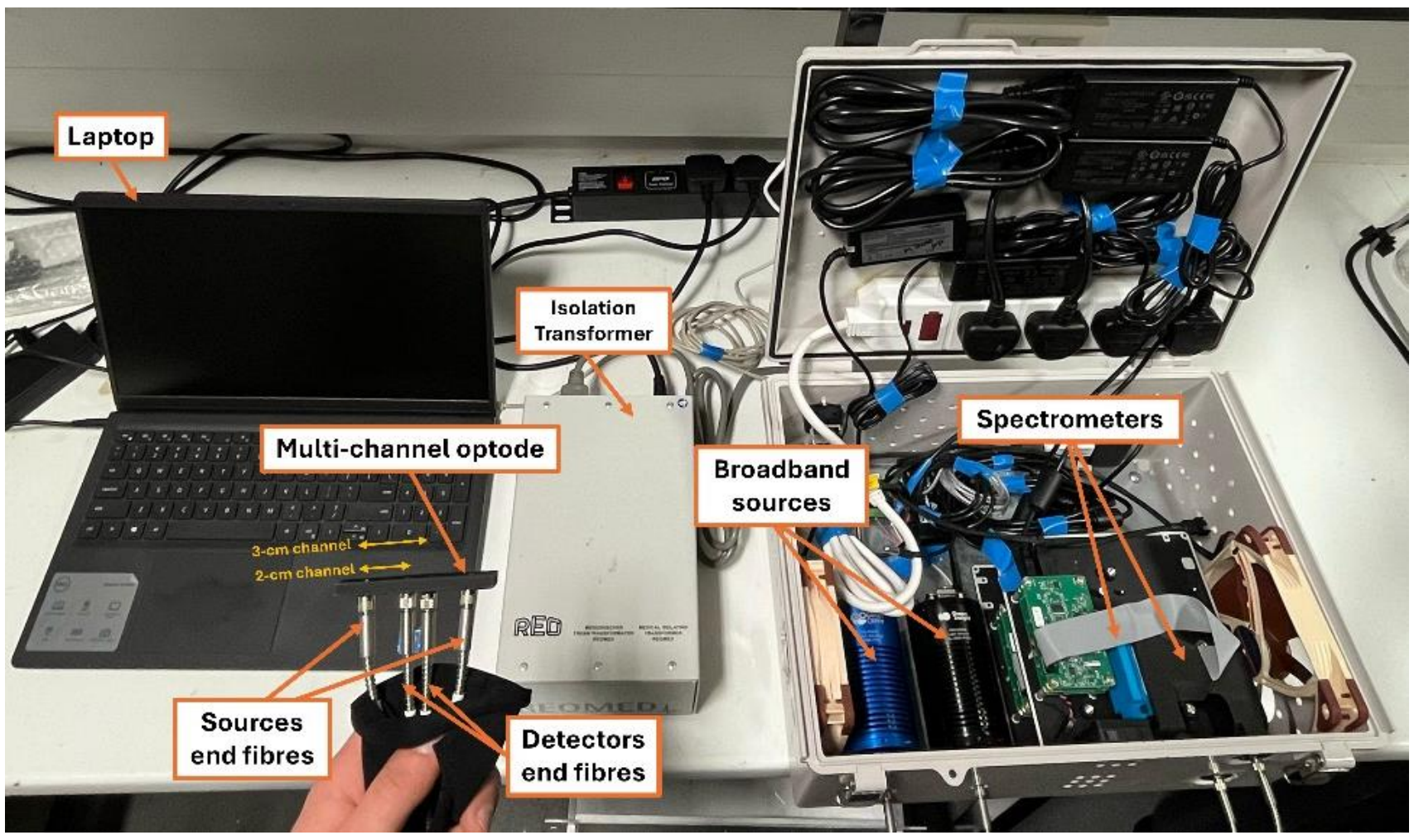

*Figure 1. Picture of the 2bNIRS system and its main components.*

2bNIRS is fully controlled via a graphic-user interface (GUI) software that provides real-time acquisition and visualisation and of the raw spectral data from bNIRS, as well as direct streamlined processing of the resulting changes in $HbO_2$, HHb, oxCCO and $StO_2$, the latter estimated with the desired multi-distance algorithm.

First, 2bNIRS was validated against a recognized "gold standard" system, the MAESTROS II developed by Lange *et al.* [8]. MAESTROS II is a time-domain NIRS (TD-NIRS) device that employs 16 wavelengths (780-870 nm, at 6-nm steps) at a 3 cm source-detector separation and estimates changes in $HbO_2$, HHb, oxCCO, as well as tissue oxygen saturation ($StO_2$) through temporal point spread function (TPSF) fitting [8]. Both systems were applied simultaneously on a dynamic, liquid optical phantom, made of water (1.4 L), Intralipid 20% (75 mL) and human blood (25 mL), reproducing the same optical contrast from haemoglobin in tissues, and with the capability of oxygenating the mixture in a controlled manner by bubbling $O_2$, and deoxygenating it via dry yeast (5 g). The latter act as a proxy for mitochondrial metabolic activity, as it contains oxCCO and consumes oxygen with the same aerobic process of mitochondria [9].

After phantom validation, two sets of use-studies on the applicability of 2bNIRS on humans were conducted: (1) one on the forearm muscle of 3 participants (2 male and 1 female, age 30±5) during arm arterial cuff occlusion (220 mmHg for 1 min), where blood flow was mechanically blocked and then released; and (2) the other on the right frontal brain cortex of 2 participants (2 male, age 35±3) during functional activation. In the latter, the participants underwent through a 60-s baseline followed by 5 blocks of 30-s task and rest, where during the task they were asked to solve mathematical equations mentally.

Algorithm-wise, 2bNIRS uses a static, wavelength-dependent differential pathlength factor (DPF), equal to 6.26 for the phantom and to 4.16 for the arm (at 780 nm), while MAESTROS II uses a temporal-dynamic, wavelength-dependent pathlength calculated directly from the time of flight of photons. Both systems then account for water fitting in the pathlength calculations.

**Results:** The validation on liquid phantom was conducted with two cycles of full oxygenation (acting as baseline for the measurements) and full deoxygenation, where the dissolved oxygen fraction ($DO_2$) in the mixture was also monitored externally and simultaneously to the 2bNIRS and MAESTROS II [9]. The results for the test are reported in Fig. 2: 2bNIRS is shown to provided consistent results in the estimates of changes in concentration of $HbO_2$, HHb and oxCCO (Fig. 2a), compared with MAESTROS II (Fig. 2b). During deoxygenation, 2bNIRS is able to capture the expected increase in HHb and decreases in $HbO_2$ and oxCCO, respectively, as well as the return to baseline of all three contributions when reoxygenation is induced, for both source-to-detector distances. Differences in the quantification of the responses between bNIRS and MAESTROS, approximately by a factor 2 for both $HbO_2$, HHb and oxCCO, could be reconducted to the differences in the calculation of the photon pathlength between the two systems.

Absolute oximetry calculated by 2bNIRS with both SRS, Dual Slope and BRUNO was also compared against the estimates from MAESTROS II (Fig. 2c), which provides the most accurate quantifications, with $StO_2$ ranging from 100% when the phantom is fully oxygenated, to nearly 0% when fully deoxygenated. Again, 2bNIRS provided consistent and comparable results with MAESTROS II, with BRUNO being the most accurate amongst the multi-distance algorithms, with a no significant difference against MAESTROS II during oxygenation and maximum difference error of ~30% during deoxygenation. Conversely, SRS provided ~10% and ~50% difference error against MAESTROS II during oxygenation and deoxygenation, respectively, whereas Dual Slope achieved ~20% and ~30% difference error in the respective cycles. Despite 2bNIRS failing into fully estimating drops to 0% with any of its oximetry algorithms, it it noteworthy that $StO_2$ levels below 30% are tipically quite rare in normal clinical routine.

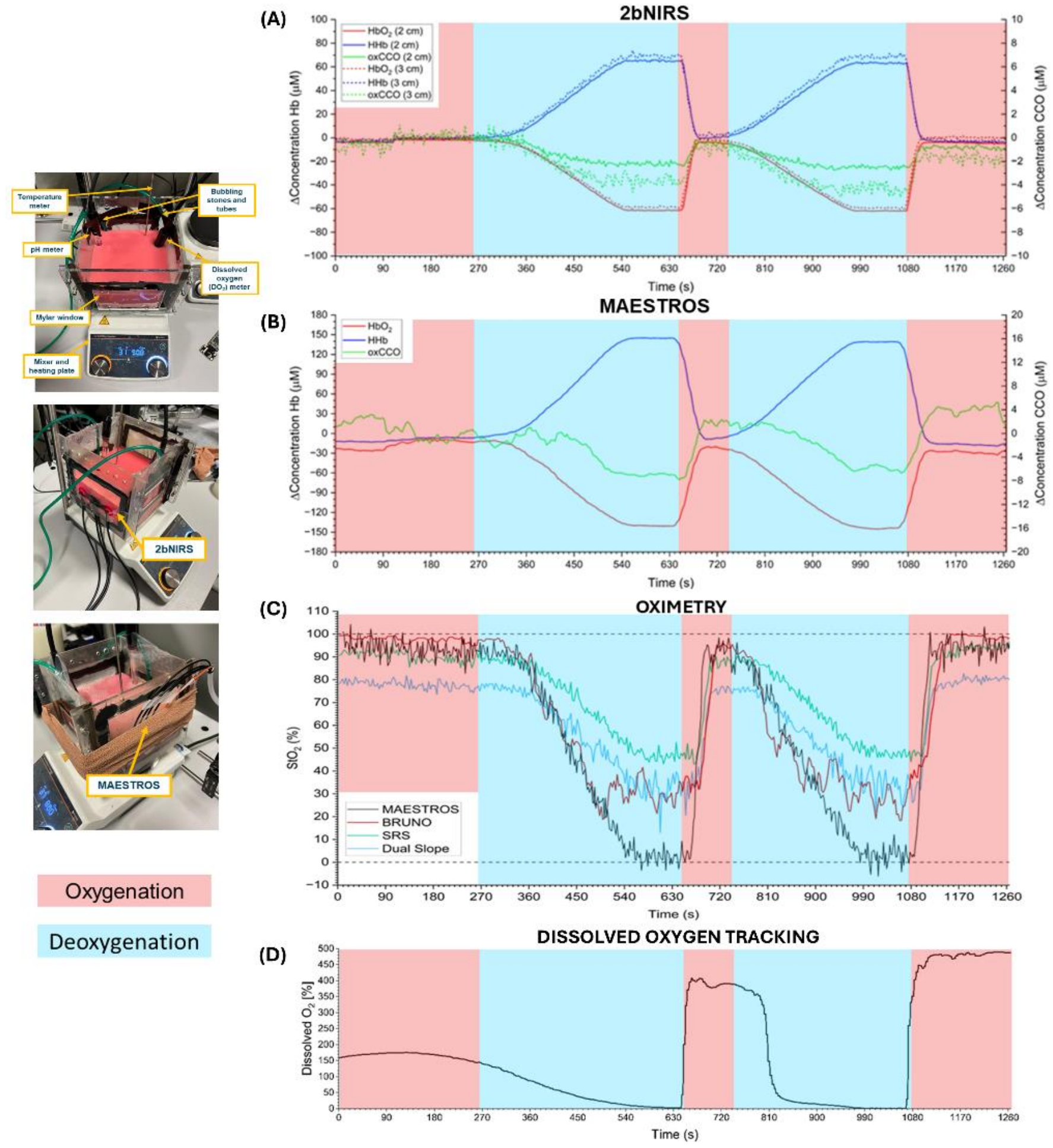

*Figure 2: Comparison of the results of the estimates of $HbO_2$, HHb and oxCCO on dynamic, blood and yeast phantom between 2bNIRS (A) and MAESTROS II (B). Comparison between the different multi-distance algorithms of 2bNIRS and TPSF fitting of absolute oximetry for MAESTROS II are also reported in (C), against the measured $DO_2$ (D).*

Fig. 3a depicts the results of the use-study with 2bNIRS on arm arterial cuff occlusion, on 3 subjects. For all the subjects, 2bNIRS successfully recovered the expected changes in $HbO_2$ and HHb, with the former decreasing and the latter increasing during the cuff, with subsequent hyperaemic effect after release (overshoot in the return to baseline of both $HbO_2$ and HHb). For all 3 subjects, 2bNIRS was also able to reconstruct the correct changes in $StO_2$ with BRUNO. In particular, for 2 out of 3 subjects, no significant changes in oxCCO were retrieved, whilst only in Subject 1 an increase in oxCCO was estimated, corresponding by far to the largest decrease in $StO_2$ (~60%, compared to 25 and 40% for Subject 2 and 3, respectively). This is in accordance with existing literature [10], suggesting that muscle metabolism during arterial occlusion should remain unchanged until all $HbO_2$ resources are completely depleted.

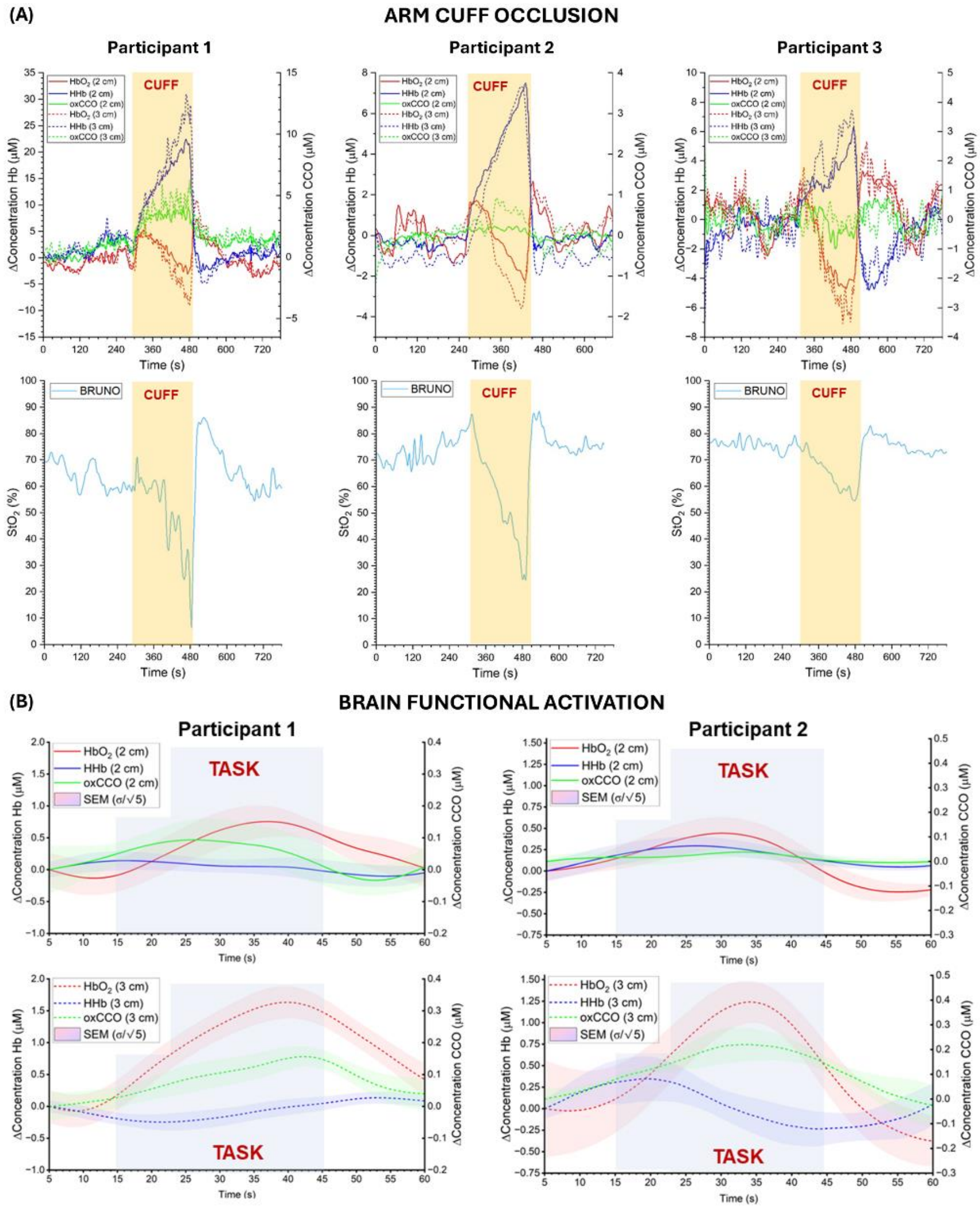

*Figure 3: Results of the arm arterial cuff occlusion experiments (A), as well as of the frontal cortex functional activation use-study (B), conducted with 2bNIRS.*

Finally, Fig. 3b presents the results of the use-study with 2bNIRS during frontal cortex functional activation on 2 subjects. The raw data were pre-processed via low-pass filtering and resampling at 1 Hz, then block averaging was performed with 60-s windows starting and ending 15 s pre- and post-task, respectively, after baseline correction (average value over the first 45 s) and normalisation to the first time point of each task window. Once again, 2bNIRS successfully reconstructed the expected physiological response of the brain during activation in both subjects, with increase in $HbO_2$ and oxCCO, and decrease in HHb. Furthermore, the comparison between the estimated changes from 2bNIRS at 2 and 3 cm provides insight on the depth sampling capability of the system in the specific scenario: at 2 cm, smaller changes in $HbO_2$ are recovered, with minimal to no changes in HHb and oxCCO, suggesting that most of the signal came from the scalp, whereas at 3 cm, the full haemodynamic and metabolic responses are estimated, indicating primary targeting of the cortex.

**Conclusion:** We introduced 2bNIRS, a novel, compact, and portable broadband NIRS system and absolute oximeter, designed to continuously measure changes in $HbO_2$, HHb, and oxidised CCO (oxCCO), as well as absolute tissue oxygen saturation ($StO_2$). This system enables real-time tracking of both haemodynamic and metabolic changes and supports multi-distance and broadband fitting spectroscopy algorithms for accurate oximetry. Housed in a transportable unit the size of a small backpack and equipped with long optical fibres for light delivery and collection, 2bNIRS is easily adaptable to any environment, including space-constrained clinical settings such as operating theatres. Its intuitive GUI ensures user-friendly operation for all types of operators.
Performance validation was conducted using a dynamic blood-and-yeast phantom replicating tissue optical properties and compared against a recognised "gold-standard", multi-wavelength TD-NIRS system. Results confirmed that the highest accuracy in tissue oximetry quantification was achieved using our BRUNO algorithm, which combines broadband fitting with spatially resolved spectroscopy (SRS), outperforming both SRS and Dual Slope (DS) methods. These findings demonstrate that integrating broadband and multi-distance spectroscopy within a continuous-wave bNIRS framework has the potential to deliver a comprehensive and versatile tool for monitoring tissue oxygen delivery and consumption.

**Bibliography:**

1. Scholkmann F, Kleiser S, Metz AJ, Zimmermann R, Mata Pavia J, Wolf U, Wolf M (2014) A review on continuous wave functional near-infrared spectroscopy and imaging instrumentation and methodology. Neuroimage 85:6–27
2. Wolf M, Ferrari M, Quaresima V (2007) Progress of near-infrared spectroscopy and topography for brain and muscle clinical applications. J Biomed Opt 12:062104
3. Matcher SJ, Kirkpatrick PJ, Nahid K, Cope M, Delpy DT (1995) Absolute quantification methods in tissue near-infrared spectroscopy. https://doi.org/101117/12209997 2389:486–495
4. Sassaroli A, Blaney G, Fantini S (2019) Dual-slope method for enhanced depth sensitivity in diffuse optical spectroscopy. J Opt Soc Am A Opt Image Sci Vis 36:1743
5. Bale G, Elwell CE, Tachtsidis I (2016) From Jöbsis to the present day: a review of clinical near-infrared spectroscopy measurements of cerebral cytochrome-c-oxidase. J Biomed Opt 21:091307
6. Kovacsova Z, Bale G, Mitra S, Lange F, Tachtsidis I (2021) Absolute quantification of cerebral tissue oxygen saturation with multidistance broadband NIRS in newborn brain. Biomed Opt Express 12:907
7. Wolf M, Greisen G (2009) Advances in Near-Infrared Spectroscopy to Study the Brain of the Preterm and Term Neonate. Clin Perinatol 36:807–834
8. Lange F, Hakim U, Highton J, Kowobari O, Ranaei-Zamani N, Johnson S, Newth O, Mitra S, Tachtsidis I (2025) MAESTROS II, a multispectral Time-Domain NIRS system for non-invasive placental monitoring. https://doi.org/10.1364/OPTICAOPEN.29941637.V1

9. Lange F, Dunne L, Hale L, Tachtsidis I (2019) MAESTROS: A Multiwavelength Time-Domain NIRS System to Monitor Changes in Oxygenation and Oxidation State of Cytochrome-C-Oxidase. IEEE Journal of Selected Topics in Quantum Electronics 25:1–12
10. Saeed F, Carter C, Kolade J, Brothers RM, Liu H (2024) Understanding metabolic responses to forearm arterial occlusion measured with two-channel broadband near-infrared spectroscopy. https://doi.org/101117/1JBO2911117001 29:117001